\documentclass[pdflatex,sn-mathphys-num]{sn-jnl}

\usepackage{graphicx}%
\usepackage{multirow}%
\usepackage{amsmath,amssymb,amsfonts}%
\usepackage{amsthm}%
\usepackage[title]{appendix}%
\usepackage{xcolor}%
\usepackage{textcomp}%
\usepackage{manyfoot}%
\usepackage{booktabs}%
\usepackage{algorithm}%
\usepackage{algorithmicx}%
\usepackage{algpseudocode}%
\usepackage{listings}%
\usepackage{enumitem}
\usepackage{physics}
\usepackage{textcomp}
\usepackage{aas_macros}
\usepackage[normalem]{ulem}  

\newcommand{\absval}[1]{\left\lvert#1\right\rvert}
\usepackage{lmodern}
\usepackage{array}

\begin{document}

\title[Article Title]{A novel technique for reflection coefficient measurement in precision cosmology}


\author*[1,2,3]{\fnm{Adarsh Kumar} \sur{Dash}}\email{akd52@cam.ac.uk}

\author[1]{\fnm{Yash} \sur{Agrawal}}

\author[1]{\fnm{Somashekar} \sur{R.}}

\author[1]{\fnm{Jayadev Ashok} \sur{T.}}

\author[1]{\fnm{Keerthipriya} \sur{S.}}

\author[1]{\fnm{Vishwapriya} \sur{Gautam}}

\author[1]{\fnm{Saurabh} \sur{Singh}}

\author[1]{\fnm{Mayuri Sathyanarayana} \sur{Rao}}

\author[1]{\fnm{Girish} \sur{B. S.}}

\author[1]{\fnm{Srivani} \sur{K.S.}}

\affil[1]{\orgname{Raman Research Institute}, \orgaddress{\street{C V Raman Avenue, Sadashivanagar}, \city{Bangalore}, \postcode{560080}, \country{India}}}

\affil*[2]{\orgdiv{Astrophysics Group}, \orgname{Cavendish Laboratory}, \orgaddress{\street{J. J. Thomson Avenue}, \city{Cambridge}, \postcode{CB3 0HE}, \country{UK}}}

\affil*[3]{\orgname{Kavli Institute for Cosmology}, \orgaddress{\street{Madingley Road}, \city{Cambridge}, \postcode{CB3 0HA}, \country{UK}}}

%

\abstract{The detection of the global 21-cm signal from the Cosmic Dawn and Epoch of Reionisation remains a challenge to experiments worldwide. Emitted at a rest-frame frequency of 1420.405~MHz, this signal from the early Universe is redshifted to 40-200~MHz with a maximum brightness temperature of a few 100~mK. Efforts to detect this sky-averaged signal include experiments such as the Shaped Antenna measurement of the background RAdio Spectrum (SARAS) and Probing ReionizATion of the Universe using Signal from Hydrogen (PRATUSH).
Detecting this faint signal requires precise calibration of the antenna, which includes a high-precision measurement of its reflection coefficient. This measurement must be performed \textit{in situ} at the observation site, as the antenna characteristics vary significantly with the environment. 
PRATUSH, a space-based radiometer, faces the additional challenge of structural distortions induced by thermal cycling, necessitating multiple measurements of the reflection coefficient. 
This work highlights the development of an \textit{in situ} Vector Network Analyser, which utilises a novel noise source-based calibration scheme and a cross-correlation spectrometer to perform magnitude and phase measurements of the complex reflection coefficient of the antenna. 
Further, we demonstrate the performance of the designed network analyser using independent measurements from a precision network analyser and reflection coefficient modelling of the device under test. We find the level of non-smooth calibration systematics, which need critical control for 21-cm signal detection, to be about $10^{-5}$. Finally, we study the impact of reflection coefficient correction on sky measurements, highlighting its usability for precision 21-cm observations.}




\maketitle

\section{Introduction}\label{Intro}
The global 21-cm signal serves as an excellent probe to study how the first stars and galaxies were formed in the period referred to as the Cosmic Dawn and Epoch of Reionisation \cite{Shaver_1999, Furlanetto_2004, Furlanetto_2006, Pritchard_2012}. Following the Cosmic Dark Ages, this period spans a redshift range of $z \sim6-35$ \cite{Cohen_Fialkov_21cm_parameter_space}. The origin of this signal is attributed to the hyperfine spin-flip transition in the abundant neutral Hydrogen (H\textsc{i}) available in the early Universe \cite{Madau_1997}. This signal was emitted at 1420.405~MHz and is expected to have been redshifted to the frequency range of 40 -- 200~MHz, with an estimated maximum absorption brightness temperature in the order of 10 -- 100~mK based on the astrophysical properties of the first sources of radiation and their interaction with the intergalactic medium \cite{Cohen_Fialkov_21cm_parameter_space}. This range, however, can be extended further in the presence of exotic models, such as the presence of an excess radio background \cite{excess_radio_edges} or a colder intergalactic medium \cite{Interstellar_gas_cooling_edges}.

Detecting this global signal has several challenges. One of them is the galactic and extra-galactic foregrounds in our targeted frequency band that can reach the orders of $10^3 - 10^4$~K in brightness temperature \cite{GSM_de_Oliveira, GMOSS_Rao_2017}. This band is also dominated by strong Radio Frequency Interference (RFI) from several terrestrial and satellite sources \cite{Offringa_RFI_LOFAR, HERA_RFI}. Various methods have been developed to mitigate the effects of such bright foregrounds and RFI, which take advantage of their different spectral behaviour compared to the global 21-cm signal. For example, given a spectrally smooth antenna and receiver response and calibration model, the measured foregrounds are expected to be largely featureless and can be fitted out using algorithms, such as maximally smooth (MS) polynomials \cite{MaxSmooth, Rao_2017}. On the other hand, the spectral features of an RFI can be narrowband or broadband, depending on its origin and can be dealt with using various flagging algorithms \cite{Bayesian_RFI_flagging, Agrawal_RFI_flagging}. 

Another challenge to the detection of the global 21-cm signal is the precise calibration of the radio telescope. These include accurate measurement or modelling of the antenna beam pattern and its reflection coefficient.
Being the first component of the signal chain, these antenna properties strongly influence the measurement of the sky spectrum \cite{Balanis_Antenna_theory}.
The environment around the antenna further affects its behaviour, especially through coupling with the surrounding medium \cite{Martha_LEDA, Agrawal_Direction_dependent_effects_2024, Pattison_environment}. 
If not corrected to the required accuracy, such calibration errors may lead to spurious detections and consequently, incorrect astrophysical interpretation \cite{Cohen_Fialkov_21cm_parameter_space}.
Moreover, the large dynamic range between the foregrounds and the 21-cm signal sets a requirement on the spectral smoothness in these measurements, which becomes significant when removing spectrally smooth foregrounds.
Therefore, measurement of these features needs to be precise and accurate, with levels better than approximately $10^{-5}$ in magnitude, to correct for any low-level features in the measured sky spectrum that could affect the detection. 
This is further explored in Monsalve et al. (2017) \cite{Monsalve_2017_EDGES_high_band_calibration}, where polynomial fits to the measured sky spectrum exhibit residuals of about 10 -- 100~mK due to errors in antenna reflection coefficient measurements. Similarly, Sun et al. (2024) \cite{sun_de_Lera_Acedo} report deviations in sky temperature measurements up to $\sim$ 10~K and propagated errors in the reconstructed 21-cm signal of approximately 200~mK and 40~mK due to errors in reflection coefficient magnitude and phase, respectively. In both works, the assumed uncertainties in the magnitude and phase were around $10^{-4}$ and $0.01^\circ$, respectively.

Various experiments follow different strategies to measure the antenna reflection coefficient for calibration. Experiments like the Experiment to Detect the Global EoR Signature (EDGES) \cite{EDGESpaper}, Radio Experiment for the Analysis of Cosmic Hydrogen (REACH) \cite{REACHpaper}, and the Mapper of the IGM Spin Temperature (MIST) \cite{MISTpaper} use an RF (Radio Frequency) switching network to toggle between science (power spectral density (PSD) measurement) mode and reflection coefficient measurement mode. 
The reflection coefficient measurements are taken using an off-the-shelf Vector Network Analyser (VNA) connected to this switching network. 
The Large-aperture Experiment to Detect the Dark Age (LEDA) does this by adding miniature coaxial test ports between the antenna and the Low Noise Amplifier (LNA) \cite{LEDApaper}. A VNA is connected to these test ports to measure the reflection coefficient at different points in the circuit. 
However, while commercial VNAs offer high accuracy and versatility, alternative approaches may offer advantages in the context of global 21-cm experiments.
Examples include lunar experiments such as PRATUSH (Probing ReionizATion of the Universe using Signal from Hydrogen) \cite{PRATUSHpaper} and CosmoCube \cite{CosmoCube_RFSoC}, which incorporate VNAs developed in-house for space-based deployment. 

The SARAS~3 experiment (Shaped Antenna measurement of the background RAdio Spectrum) \cite{SARAS3paper} uses a floating disk-cone antenna over water bodies to avoid multi-path reflections arising due to a stratified ground underneath the antenna. The antenna reflection coefficient measurements are performed using a replica of the front-end receiver box, housing a commercially available VNA \cite{SARAS_floating_paper}. 
This method, however, makes it infeasible to repeat these measurements over water due to practical challenges associated with deployment, including surrounding topography, weather and related environmental factors in remote areas. Further, the instability of the antenna and its environment in water-based deployments advocates an \textit{in situ} VNA unit in this case \cite{Agrawal_Direction_dependent_effects_2024}.
Similarly, for space-based experiments, the antenna can undergo further warping or exhibit structural deformities due to thermal cycling in its orbit, thereby increasing the necessity for a system to perform repeated reflection coefficient calibration and correction over the mission period.

Early network analysers, developed during the 1950s and 1960s, incorporated several advancements in RF engineering, such as swept-frequency sources like the Backward Wave Oscillators, frequency mixing techniques, and complex reflection and transmission measurement using broadband sampling techniques. 
The metrology of VNAs continues to be an active area of research. Significant efforts are focused on improved calibration techniques, characterisation of non-ideal standards, verification methodologies, and uncertainty estimation \cite{TRL, Traceability_Wong, Callegaro_2009, EURAMET_Calibration_Guide, One_port_systematics_poorL, Nport_error_model}.
In parallel, the introduction of the Smith Chart by Smith (1939) \cite{Smith1939} provided a useful graphical tool to interpret reflection coefficient and impedance measurements, and is widely used today.
A historical overview of the major developments and milestones can be found in Rytting (2008) \cite{Rytting_VNA_history} and Bernardi and Marasca (2025) \cite{Bernardi_VNA_milestones}.

Motivated by these developments, this work presents a noise-source-based VNA (NSVNA) architecture and investigates the challenges and advantages associated with its design.
The proposed architecture builds upon established VNA principles, while introducing a noise source for signal generation and a cross-correlation spectrometer, both of which have been explored previously in other measurement contexts, their combination in VNA design has not been extensively investigated.
We present the NSVNA as a calibration approach designed for integration with the receivers of various global 21-cm experiments, in particular, SARAS and PRATUSH.
This class of experiments are very sensitive to potential spectral distortions due to systematics, which may introduce spectral features and compromise signal detection.
We begin by evaluating the NSVNA performance for the required accuracy using measurements from a high-accuracy commercial VNA as a first-level benchmark. 
We focus specifically on one aspect of measurement quality, namely non-smooth calibration systematics, and study its implications for 21-cm signal detection.
To this end, we employ a custom-made Resistor-Inductor-Capacitor network (RLC) that mimics the reflection coefficient of the SARAS antenna.
This network is first used to compare the level of non-smooth calibration systematics in the NSVNA and PNA measurements. We then perform an illustrative test on how these systematics propagate into an idealised simulated 21-cm signal detection scenario in the context of SARAS measurements.

The paper begins by defining the basic terms used in the later sections in the context of VNA design in Section~\ref{Sec:Basic_def}. This is followed by Section~\ref{Sec:SARAS_RL_measurement}, where we discuss the current method of reflection coefficient measurement of the SARAS antenna. The discussion also motivates the need for an \textit{in situ} setup. 
Section~\ref{Sec:Methodology} explains the design of the developed NSVNA, its data acquisition methods, as well as the various considerations that led to the final design. 
The acquired data from the NSVNA then need to undergo an offline calibration process, which is detailed in Section~\ref{Sec:offline_calibration}. The results after the offline calibration are also presented in this section, ending with a discussion on its accuracy.
Section~\ref{Sec:result_validation} focuses on the various validation techniques that we used for the results, ending with comments on the impact on sky measurements that could affect the 21-cm signal detection prospects. We conclude the paper by summarising and discussing the future of this work in Section~\ref{Sec:Summary}.

\section{Basic definitions} \label{Sec:Basic_def}

The reflection coefficient ($\Gamma$) of a device at a given port is the complex ratio of voltage waves travelling away from the port $(V^-)$ to those travelling towards it $(V^+)$. In addition to the above definition, it can also be defined using a load impedance ($Z_L$) and the characteristic impedance of the transmission line ($Z_0$) as
\begin{equation} \label{eq:reflection_coeff_def}
    \Gamma = \frac{V^-}{V^+} = \frac{Z_L - Z_0}{Z_L + Z_0},
\end{equation}
where each term is a function of frequency \cite{Pozar:882338}.

\begin{figure}[!htbp]
    \centering
    \includegraphics[width=0.7\linewidth]{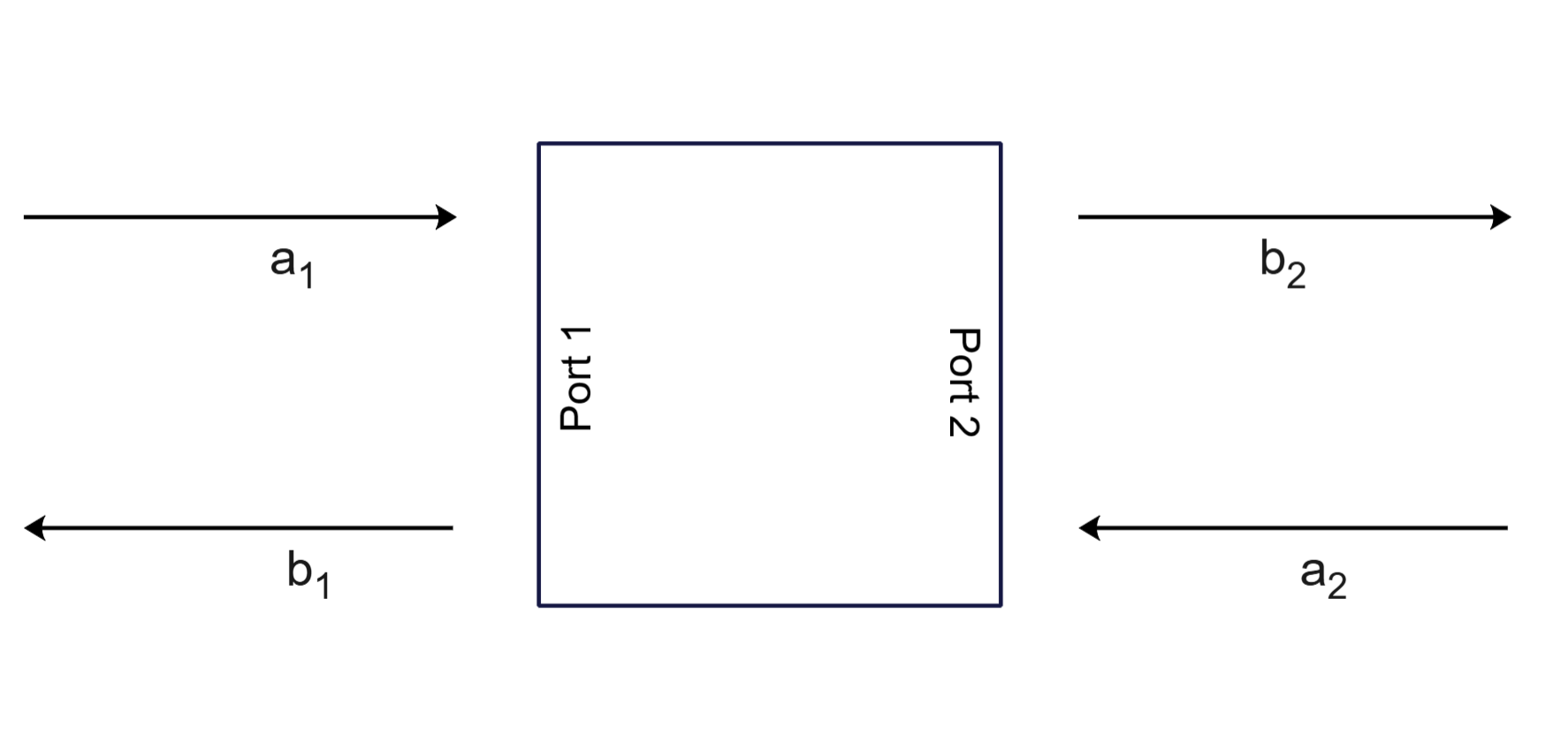}
    \caption{A simple 2-port network is shown. Here, $a_1$ and $a_2$ are the incident waves, and $b_1$ and $b_2$ are the outgoing waves.}
    \label{fig:2port_network}
\end{figure}

Another important term used in this work is scattering parameters (s-parameters), defined using voltage travelling waves. Any electrical network that is assumed to be linear and time-invariant can be described using a set of s-parameters as a function of frequency for a given characteristic impedance ($Z_0$) \cite{Pozar:882338}. 
S-parameters can be explained by taking a simple example of a 2-port network, as shown in Figure~\ref{fig:2port_network}. The incident waves are shown as $a_1$ and $a_2$, while $b_1$ and $b_2$  are termed the outgoing waves. The S-parameter matrix for a 2-port network is defined using a set of complex numbers as
\begin{equation} \label{eq:2port_s_matrix}
    \begin{pmatrix}     b_1 \\ b_2      \end{pmatrix} 
= 
\begin{pmatrix}     S_{11} & S_{12} \\ S_{21} & S_{22}      \end{pmatrix} 
\begin{pmatrix}     a_1 \\ a_2      \end{pmatrix}.
\end{equation}

The amplitudes of all voltage waves are measured with the ports terminated with a matched load, which has an impedance equal to the characteristic impedance, $Z_0$ (usually $50~\Omega$). 
For example, if Port 2 of the network in Figure~\ref{fig:2port_network} is terminated with a matched load, following Equation~\ref{eq:reflection_coeff_def}, $b_2$ is completely absorbed by the load and $a_2$ becomes zero. Solving the matrix in Equation~\ref{eq:2port_s_matrix} we get,
\begin{equation}
    S_{11} = \frac{b_1}{a_1} \hspace{1cm} \text{and} \hspace{1cm} S_{21} = \frac{b_2}{a_1}.
\end{equation}
Similarly, terminating Port 1 with a matched load gives us
\begin{equation}
    S_{22} = \frac{b_2}{a_2} \hspace{1cm} \text{and} \hspace{1cm} S_{12} = \frac{b_1}{a_2}.
\end{equation}
The terms $S_{11}$ and $S_{22}$ are defined as input and output reflection coefficients, while $S_{21}$ and $S_{12}$ are the forward and reverse voltage gain (or loss), respectively. For a one-port device, we consider only its input reflection coefficient ($S_{11}$).

As an equivalent quantity to the reflection coefficient, Return Loss (RL) is sometimes used, measuring the loss in the returned power ($P_r$) with respect to the incident power ($P_i$) on a dB scale, due to an impedance mismatch. It is defined to be a positive quantity. The set of equations below summarises the relation between RL, $\Gamma$, and $S_{11}$ for a one-port device.
\begin{equation} \label{eq:s11gamma_RL_relation}
\begin{split}
    \mathrm{RL} (\mathrm{dB}) &= 10 log_{10} \frac{P_i}{P_r} 
    = - 20 log_{10} \frac{\absval{V_r}}{\absval{V_i}} ,\\
    \mathrm{RL} (\mathrm{dB}) &= - 20 log_{10} \lvert\Gamma\rvert = -20 log_{10}\lvert S_{11}\rvert ,
\end{split}
\end{equation}
where we have used the fact that power $P_i$ ($P_r$) is proportional to square of the voltage $V_i$ ($V_r$). Because of their simple relation in Equation~\ref{eq:s11gamma_RL_relation}, these terms may be used interchangeably throughout the paper.

Moving to the context of multi-port devices, we use the definition of Insertion Loss (IL) as the loss in signal power when a device is inserted in a transmission line and is given by the reciprocal of the gain.
Another term used here is the isolation between two ports. For amplifiers, the corresponding quantity of interest is the reverse isolation $I_\mathrm{rev}$. Both of these quantities are given below.
\begin{equation}
\begin{split}
     \mathrm{IL} (\mathrm{dB}) &= -20 log_{10} \absval{S_{21}} ,
    \\
    I_\mathrm{rev} (\mathrm{dB}) &= \absval{20 log_{10} \absval{S_{12}}} .
\end{split}
\end{equation}

For directional couplers, we define multiple quantities (refer to Figure~\ref{fig:vna_simple}). First is the isolation between the coupled and the output port, and we call it the isolation of a coupler, $I_\mathrm{cpl}$.
Next is the forward Coupling Factor (CF), defined as the ratio of powers at the input port to that at the coupled port, in dB scale.
Lastly, we use the definition of directivity for a coupler as its ability to separate signals moving forward and backwards through it. Mathematically,  in dB scale,
\begin{equation} \label{eq:Directivity_def}
\mathrm{Directivity} = I_\mathrm{cpl} -
\mathrm{CF} - \mathrm{IL},       
\end{equation}
where IL is the insertion loss between the output and input ports.

\section{The SARAS~3 case}\label{Sec:SARAS_RL_measurement}

The SARAS~3 experiment uses a floating disk-cone antenna and front-end receiver, with its back-end receiver and digital correlator placed at the shore, separated by a 150~m optical fibre \cite{SARAS_floating_paper}. The front-end receiver is equipped with Li-ion batteries that can power the system without requiring any external power supply. 
The separation helps mitigate measurement systematics by providing excellent galvanic and reverse isolation between the front-end and back-end electronics. Standing waves introduced downstream of this stage are expected to be calibrated out as part of Dicke Switching and differencing methods \cite{SARAS3paper}.
Ground-based experiments with a small ground plane, such as SARAS, also face the challenge of modelling multi-path reflections from the ground, which arise due to stratified soil structure. The challenge is commonly mitigated by deploying the antenna over a conductive ground plane, providing a more controlled and reliable electromagnetic environment. The floating SARAS radiometer follows a similar philosophy. The water medium reduces these reflections and can be modelled using fewer parameters compared to a multi-layered soil. 
In addition, the antenna is designed to have spectrally smooth $S_{11}$ over such media to avoid any spurious signatures.
Nevertheless, factors such as the orientation of the antenna-front-end raft, water conductivity and water depth can influence the antenna response \cite{Agrawal_Direction_dependent_effects_2024}.
To keep our science instrument as compact and lightweight as possible, the antenna $S_{11}$ measurement is carried out using a separate receiver unit, which houses a FieldFox VNA \cite{Keysight_datasheet}, hereafter referred to as the FieldFox $S_{11}$ unit. 
This unit is a well-shielded, waterproof aluminium enclosure and is a replica of the main front-end receiver box with similar electromagnetic properties. It is connected to the `shore electronics' using a 150~m optical fibre, while being deployed at the same position where the main receiver is placed on water (or land) to minimise the effects due to a change in surroundings.

The VNA \textit{in situ} calibration and antenna $S_{11}$ measurement are performed using a 4-way electro-mechanical switch, which has a low insertion loss and high isolation between its ports.
The VNA inside the FieldFox $S_{11}$ unit is connected to the common port of this switch using a short-length cable.
Three ports of the switch are connected to precision standard terminations (Open, Short, and $50~\Omega$ broadband load) with known RL characteristics, and the last port is connected to the Device Under Test (DUT), which in this case is the antenna. 

Data are recorded while switching between the three standards and the DUT, where each cycle constitutes the states O-S-L-D-O-S-L.
Each state is observed for 800 iterations ($\sim$ 7~min) before switching to the next. Ideally, switching would occur at every iteration. However, the practical limitation of the switch lifetime necessitates a longer dwell time. This choice also balances the signal-to-noise ratio against drift in the system gain over extended observation periods.
This whole process is repeated multiple times to further correct for any gain drift errors. Each cycle takes a total time of about 45 minutes. The total measurement time is limited by the battery capacity of the VNA, which leads to around 3 cycles in total.

\subsection{The limitations}

The SARAS~3 $S_{11}$ measurement strategy has a few limitations. First is the challenge of deployment and operation in a remote water-body environment, where local topography, weather conditions, site accessibility, and logistical constraints can significantly complicate the installation of the radiometer.
These challenges limit the repeatability of reflection coefficient measurements when interchanging the front-end receiver unit with the FieldFox $S_{11}$ unit. 
Secondly, the high power consumption of the VNA limits its operational period, thereby increasing the statistical uncertainty of the $S_{11}$ measurement. 
Lastly, compared to ground-based methods, a water-based deployment is prone to more variation in the antenna location, orientation, tilt and raft height, as well as changes in the conductivity of the water medium. These can severely affect the antenna beam pattern and $S_{11}$, introducing systematic errors and, in turn, degrading 21-cm signal detection prospects \cite{Agrawal_Direction_dependent_effects_2024}.

\begin{figure}[!htbp]
    \centering
    \includegraphics[width=0.95\linewidth]{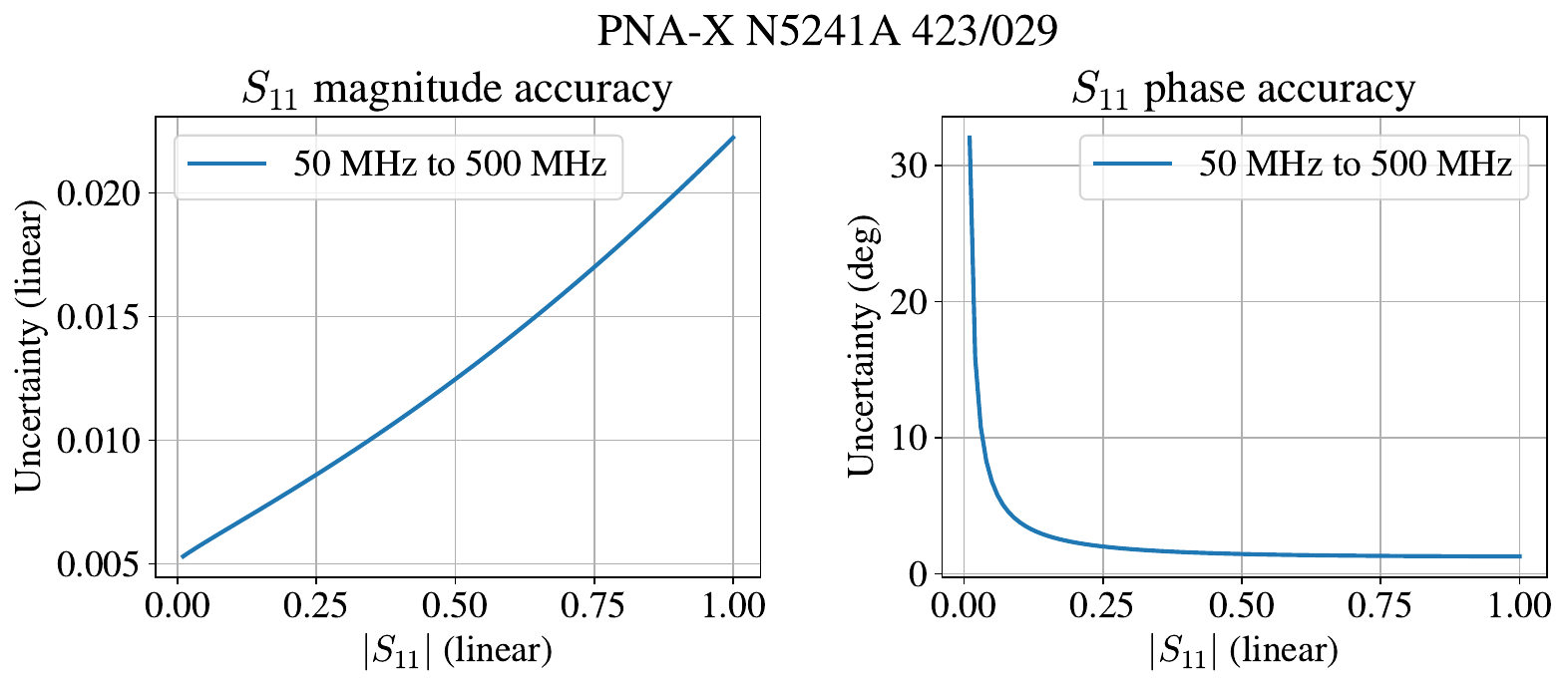}
    \caption{Dependence of $S_{11}$ uncertainty for the PNA-X N5241A 423/029 VNA in the frequency range of 50-500~MHz, obtained using the VNA uncertainty calculator provided by Keysight Technologies \cite{uncert_calculator}. The calibration and measurement powers were chosen to be -5~dBm, using the Ecal module N4691B as the calibration kit. The averaging factor was 200.}
    \label{fig:PNA_calculated_uncert}
\end{figure}

When considering off-the-shelf VNAs, it is important to note that they provide excellent performance across a wide range of applications. 
For a typical VNA, the specified measurement uncertainty is of the order of 0.1~dB or $10^{-2}$ in magnitude and $1^\circ$ in phase for frequencies below 200 MHz.
This can be observed in Figure~\ref{fig:PNA_calculated_uncert}, generated using the VNA uncertainty calculator provided by Keysight Technologies \cite{uncert_calculator}, taking the PNA-X N5241A 423/029 VNA as an example. 
The achievable performance depends on several factors, including the calibration kit, connector repeatability, frequency range, and environmental stability \cite{sun_de_Lera_Acedo}. 
While commercial VNAs are well-suited for a broad range of measurement scenarios, the exceptionally stringent and unique requirements of 21-cm cosmology motivate alternative design approaches, tailored to the problem of global 21-cm radio cosmology.
In this work, we focus on a specific characteristic of $S_{11}$ measurement, namely its spectral smoothness. While we assume the PNA as a benchmark for accuracy, we argue that such custom designs and smoothness criteria might prove helpful in the context of our science case.

\section{Development of noise source-based VNA}\label{Sec:Methodology}

\begin{figure}[!htbp]
    \centering
    \includegraphics[width=1\linewidth]{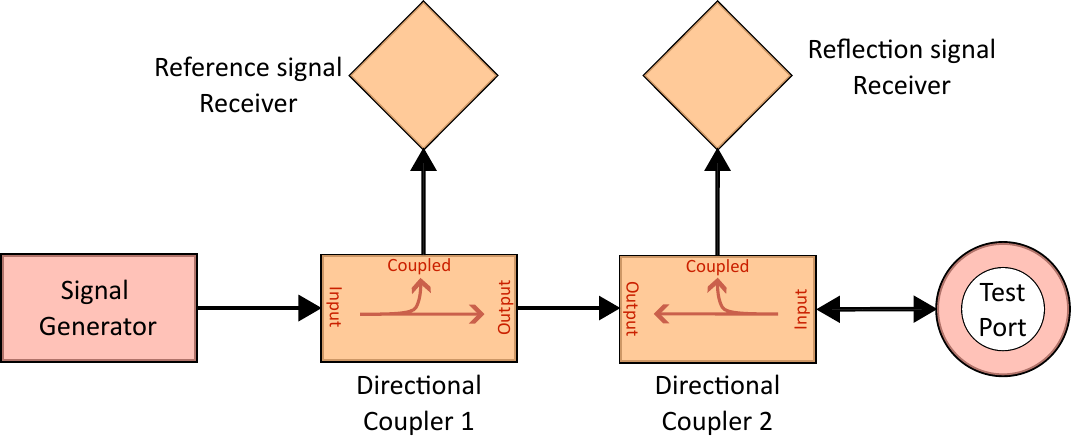}
    \caption{An illustration of a simple network analyser setup. The black arrows indicate the direction of signal flow. Directional Coupler~1 is used to get a reference signal. Directional Coupler 2, as indicated by the red arrows, is operated in a reversed mode and is used to sample the reflected power from the DUT at the test port. Both powers are measured by the corresponding receivers.}
    \label{fig:vna_simple}
\end{figure}

The reflection coefficient of an RF component is primarily measured using a network analyser. Depending on the application, the measurement can be scalar, providing only the magnitude, or vector, providing both magnitude and phase of the reflection coefficient.
A typical VNA achieves this using a signal generator, directional couplers, and dedicated receivers for reference and reflected signals. The simultaneous reference measurement allows correcting output fluctuations in the signal generator and common-mode receiver gain variations. It also enables phase measurement of the reflected signal. The signal flow is illustrated in Figure~\ref{fig:vna_simple}.
The DUT is connected to the test port during a measurement. Similarly, to calibrate the VNA, known calibration standards are sequentially connected to the test port, which are used to determine the systematic error of the measurement system \cite{VNA_error_calibration}.

As the NSVNA is intended to be integrated with the SARAS and PRATUSH receiver, its design adopts several design elements from the existing SARAS receiver architecture.
In particular, the receiver blocks shown in Figure~\ref{fig:vna_simple} are implemented using the SARAS cross-correlation spectrometer, with a frequency resolution of 61~kHz \cite{SARAS_digital}. This enables the measurement of both magnitude and phase of the incoming signals.
The following sections elaborate on the choice of signal generator and the design of the conditioning circuit, which enable the \textit{in situ} calibration of the developed VNA and the measurement of the reflection coefficient.

\subsection{Signal generator}
The type of signal generators used usually depends on the DUT characteristics, the receiver specifications and the mode of data acquisition. A standard choice can be tone-based signal generators. For example, a LimeSDR \cite{LimeSDR_web}, having its own Tx and Rx ports, is suitable for use in sweep mode. Similarly, a frequency comb generator can be used, which takes an input frequency and generates its harmonics in the form of a comb of frequency lines. These sources have high signal-to-noise ratio; however, high spectral resolution usually comes at the expense of longer sweep times.  

Another method of signal generation is a broadband noise source. In this case, the resolution is limited by the receiver in use. 
A noise-source signal can be sampled directly in the band of interest, which is enabled by the SARAS spectrometer \cite{SARAS_digital}.
However, unlike a tone generator, the available source power is distributed across the entire bandwidth. As a result, the total broadband power is limited by the full-scale input of the ADC, thus reducing the maximum usable power per frequency channel. This limit is approximately -2~dBm for the currently used spectrometer. At the lower end, the quantisation noise power referred to the ADC input is $\sim$~-55~dBm.

This work explores the noise source-based signal generation. We use a broadband noise source model NC1113B (Noisecom). The datasheet provides specifications mentioning a flatness of $\sim$~0.04~dB in the frequency band of interest and typical temperature stability of 0.025~dB$/^{\circ}\mathrm{C}$ \cite{noise_source_datasheet}. 
A flat response is important to maximise the use of the receiver's dynamic range, spanning between reflected powers from Open/Short and the $50~\Omega$ termination states. The higher temperature stability further reduces the drift errors in the system.

\subsection{Measurement setup} \label{subsec:Measurement_setup}

The output power of the signal generator was adjusted based on the linear range of operation of the SARAS digital correlator, which acted as the digital receiver for $S_{11}$ measurements. The total power received by the correlator varies as the switch cycles through Open, Short and $50~\Omega$ (referred to as the Load) terminations. Naturally, the highest power is received when connected to Open and Short terminations (total $\approx$~-10.5~dBm), while the Load termination results in the lowest power (total $\approx$~-49~dBm) due to a negligible reflection coefficient. The total reference power was measured close to -38~dBm.

\begin{figure}[!htbp]
    \centering
    \includegraphics[width=\textwidth]{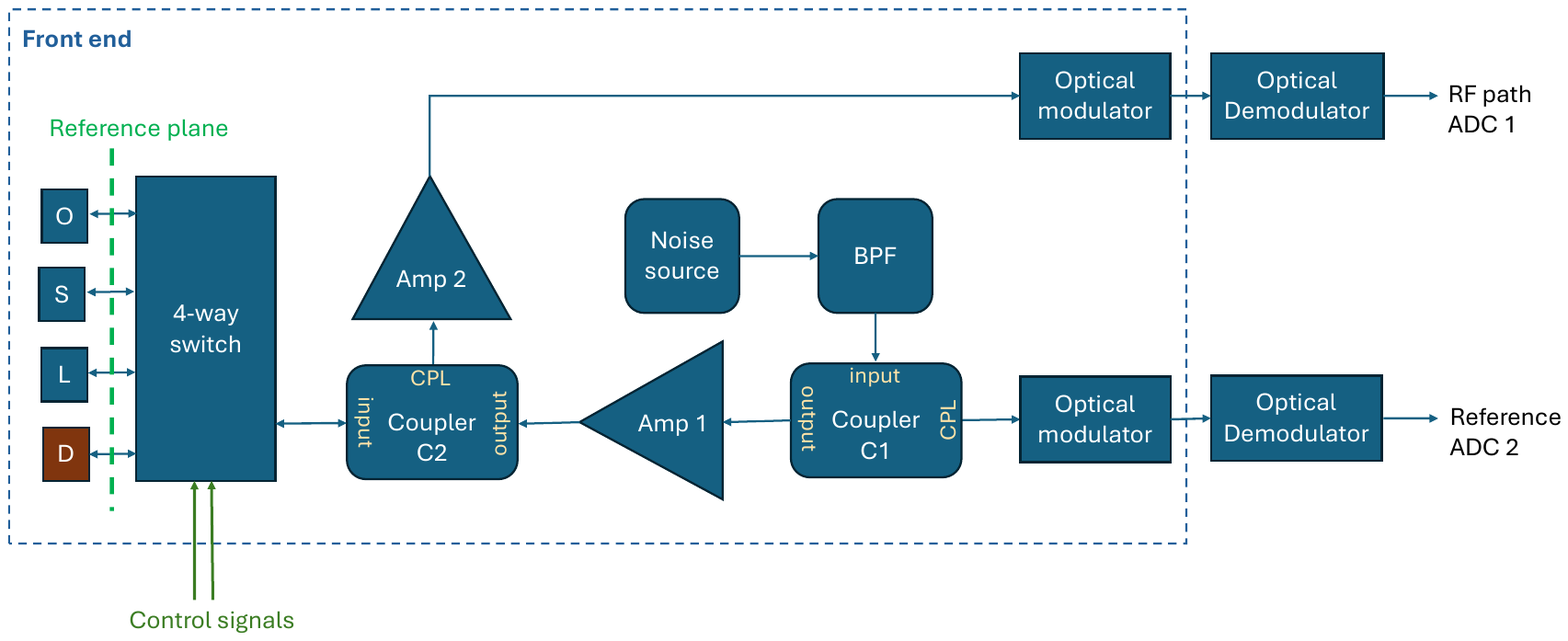}
    \includegraphics[width=0.9\textwidth]{Images/vna_photo_labelled.pdf}
    \caption{\textit{Top:} A detailed block diagram of the NSVNA. The arrows indicate the direction of signal flow. Attenuators were added to tune the power at the ADCs of the spectrometer and improve the matching of filters and amplifiers, which are not shown in the diagram for visual simplicity. 
    Connected to the 4-way switch are the three standard terminations Open (O), Short (S) and Load (L), as well as the DUT (D). These four ports represent the calibration reference plane of the VNA.
    \textit{Bottom:} A top-view photograph of the NSVNA. Some of the components are not visible, as they are positioned below another component. The standard terminations, DUT and the optical modulators have not been connected.}
    \label{fig:s11_block}
\end{figure}

The final setup is shown in Figure~\ref{fig:s11_block} and is constructed as follows: 
\begin{enumerate}
    \item The noise source is followed by a bandpass filter (40 - 110~MHz) and a series of attenuators, resulting in a `conditioned signal' (total power $\approx$~-26~dBm). This conditioning keeps the total power input within the desired dynamic range.
    \item The conditioned signal is fed to a directional coupler (C1), where the output path is labelled as the `RF path', and the coupled path is used as a `reference path' for both magnitude and phase measurements.
    \item The reference path is further electrically isolated via an optical modulator.
    \item The RF path consists of amplifiers, a second directional coupler (C2) and a 4-way switch.
    \item Post-amplification at Amp~1, the total noise power is approximately -~10~dBm. The injected signal is reflected by one of the terminations connected via a 4-way switch. The switch has been added to perform multiple calibration cycles of the VNA over the observation period and can toggle among the four states: three precision standards and the fourth acting as the test port connected to the DUT.
    \item The reflected signal is sampled by the coupled output of C2. Then, it is amplified at Amp~2 and isolated optically.
    \item The two output signal chains undergo optical demodulation and are later connected to the ADC inputs 1 and 2 of the SARAS correlation spectrometer.
\end{enumerate}

The switching between the four measurement states is controlled by an Arduino UNO. The whole setup is powered by Li-ion batteries.
This architecture provides high reverse and galvanic isolation of the $S_{11}$ unit from the digital correlator. Additionally, it makes it suitable for integration with the SARAS and PRATUSH deployment setup \cite{PRATUSH_SBC_digital}.
The final version of the NSVNA included numerous iterations and considerations. These are summarised in Table~\ref{tab:Additional_considerations} along with the specifications of the key components.
Subsequent offline calibration is performed to obtain the final $S_{11}$ of the DUT, which is explained in Section~\ref{Sec:offline_calibration}.

\begin{table}
    \centering
    \caption{A summary of the key specifications of components used in the NSVNA with their purpose. For most components, the typical values within the band of interest are shown as taken from their datasheets. The isolation of couplers is calculated using Equation~\ref{eq:Directivity_def}. The directivity and source matching errors are defined later in Section~\ref{sec:process_and_formalism}.}
    \begin{tabular}{|m{0.25\textwidth}|m{0.18\textwidth}|m{0.14\textwidth}|m{0.3\textwidth}|}
        \hline
        Component & Key specification & Value & Purpose \\
        \hline \hline
        Noise source: {NC1113B} & Power output & -95 dBm/Hz &  Power input at the VNA test port. \\
            & Flatness & 0.04 dB & Maximise the use of dynamic range. \\
        \hline
        C1: {ZFDC-20-4L} \cite{ZFDC_20_4L_datasheet} & Isolation & 60.4 dB & Isolation of the RF and reference paths. \\
        \hline
        Amplifier~1: {WHM0012AE} \cite{WHM0012AE_datasheet} & Noise Figure & 1.25 dB & Low noise in the initial stage of the RF path. \\
            & Gain & 32 dB & Power input at the VNA test port. \\
            & $I_\mathrm{rev}$ & 45 dB & Isolation of the RF and reference. \\
        \hline
        C2: {ZFDC-20-3+} \cite{ZFDC_20_3_datasheet} & Directivity & 36 dB & Reduces the directivity error. \\
            & Output RL & 34.10 dB & Reduces source matching error. \\
            & Isolation & 55.85 dB & Isolation between the input signal and reflected signal coupled to C2. \\
        \hline
        Amplifier~2: {QB-300} \cite{QB_300_datasheet} & Gain & 24.5 dB & Optimal power at later stages and ADC. \\
        \hline
    \end{tabular}
    \label{tab:Additional_considerations}
\end{table}

\section{Offline calibration: Methodology and Results}\label{Sec:offline_calibration}

\subsection{The process and calibration error formalism} \label{sec:process_and_formalism}

The calibration process is performed using the 85052D 3.5~mm Calibration Kit \cite{CalKit_datasheet}. A single calibration measurement consists of cycling through Open-Short-Load, followed by the DUT, and another cycle of Open-Short-Load. In total, we therefore cycle through seven system states. The measurements are taken for 10 minutes for each state. This rate was chosen to provide an optimal period such that the system maintains stability during the observation with $\sim~400$ traces per state. This rate also helped preserve the operational lifetime of the switch. The traces for each state are then averaged to reduce random noise.  

\begin{figure}[!htbp]
    \centering
    \includegraphics[width=0.48\linewidth]{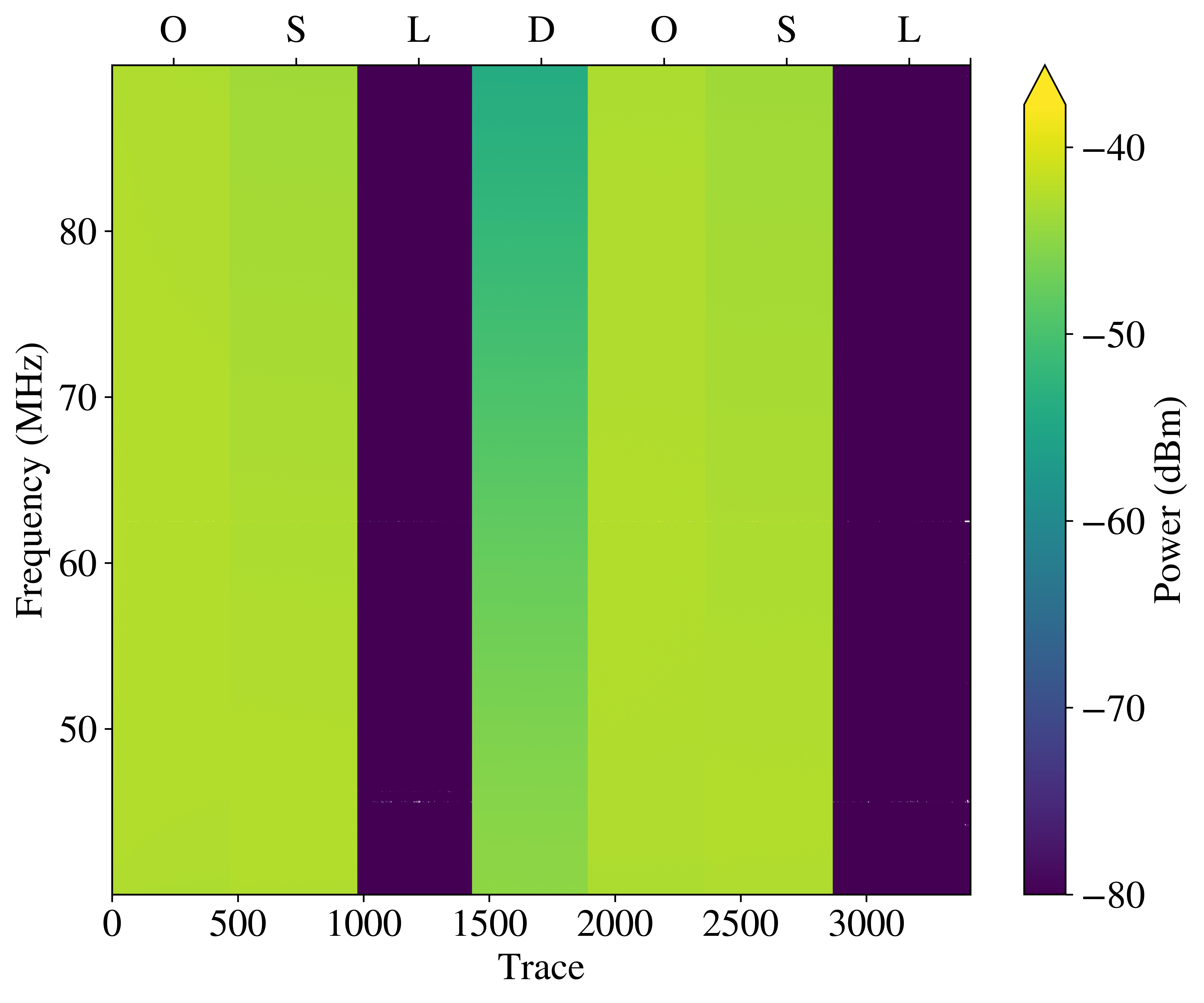}
    \includegraphics[width=0.48\linewidth]{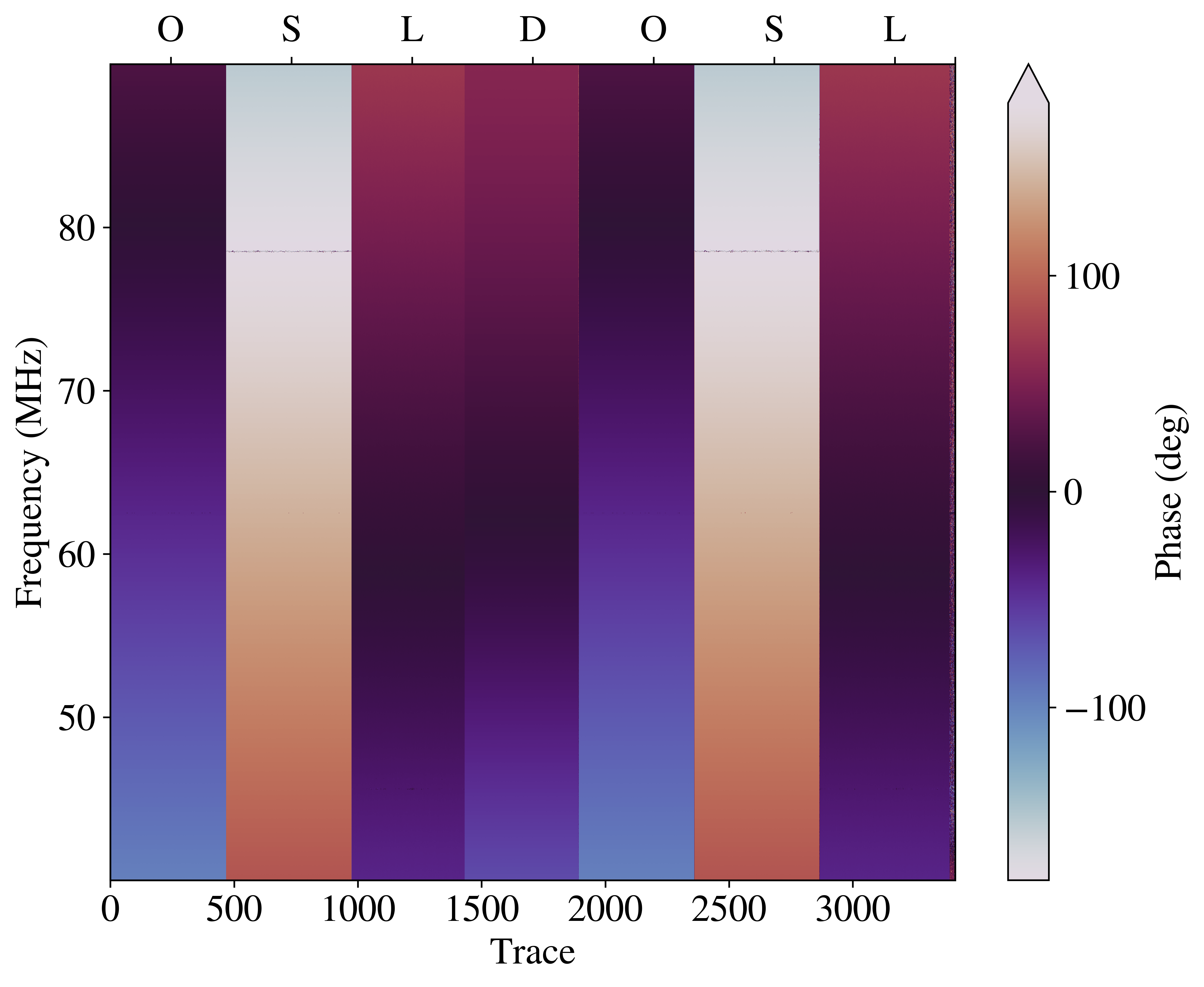}
    \includegraphics[width=0.49\linewidth]{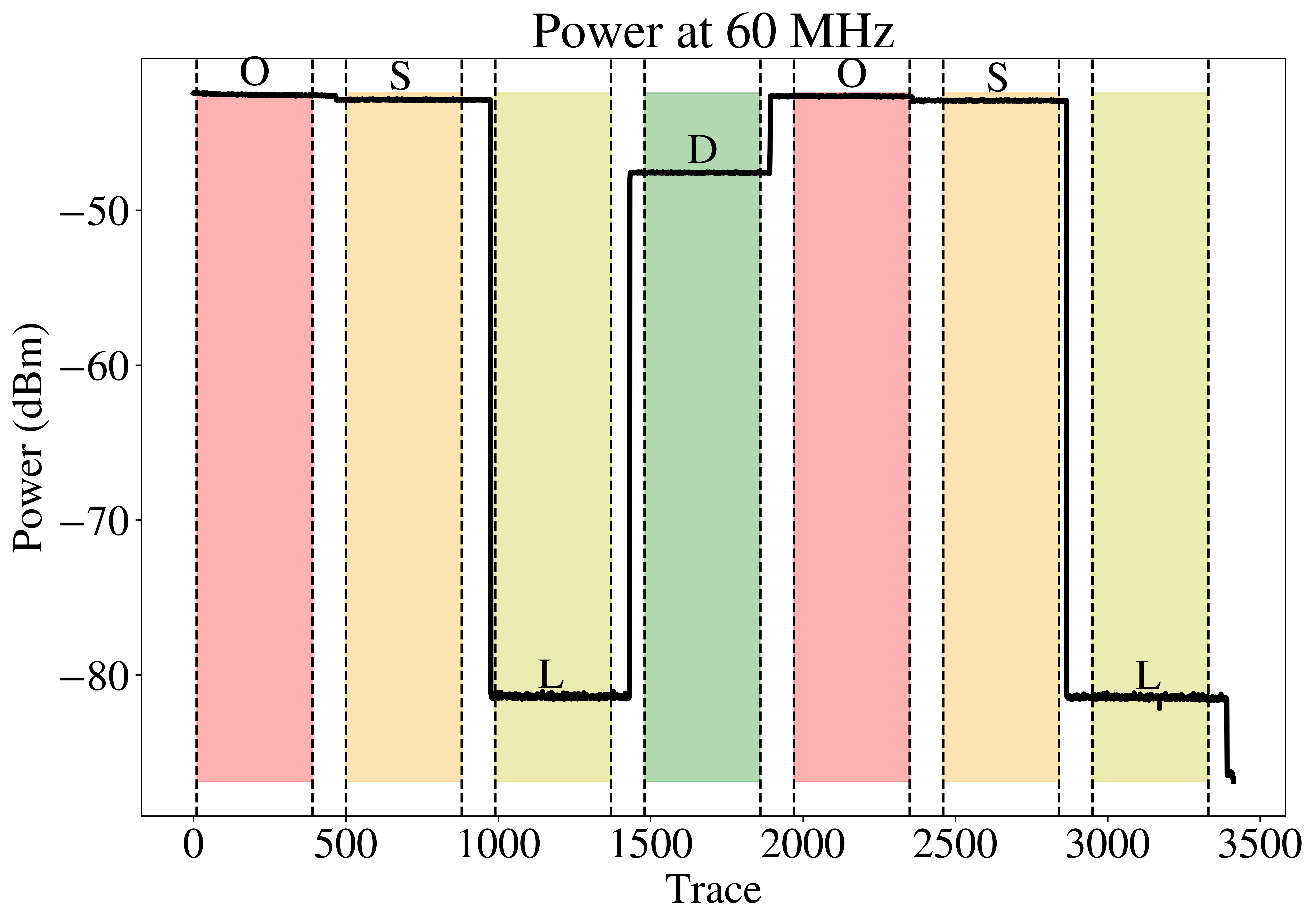}
    \includegraphics[width=0.49\linewidth]{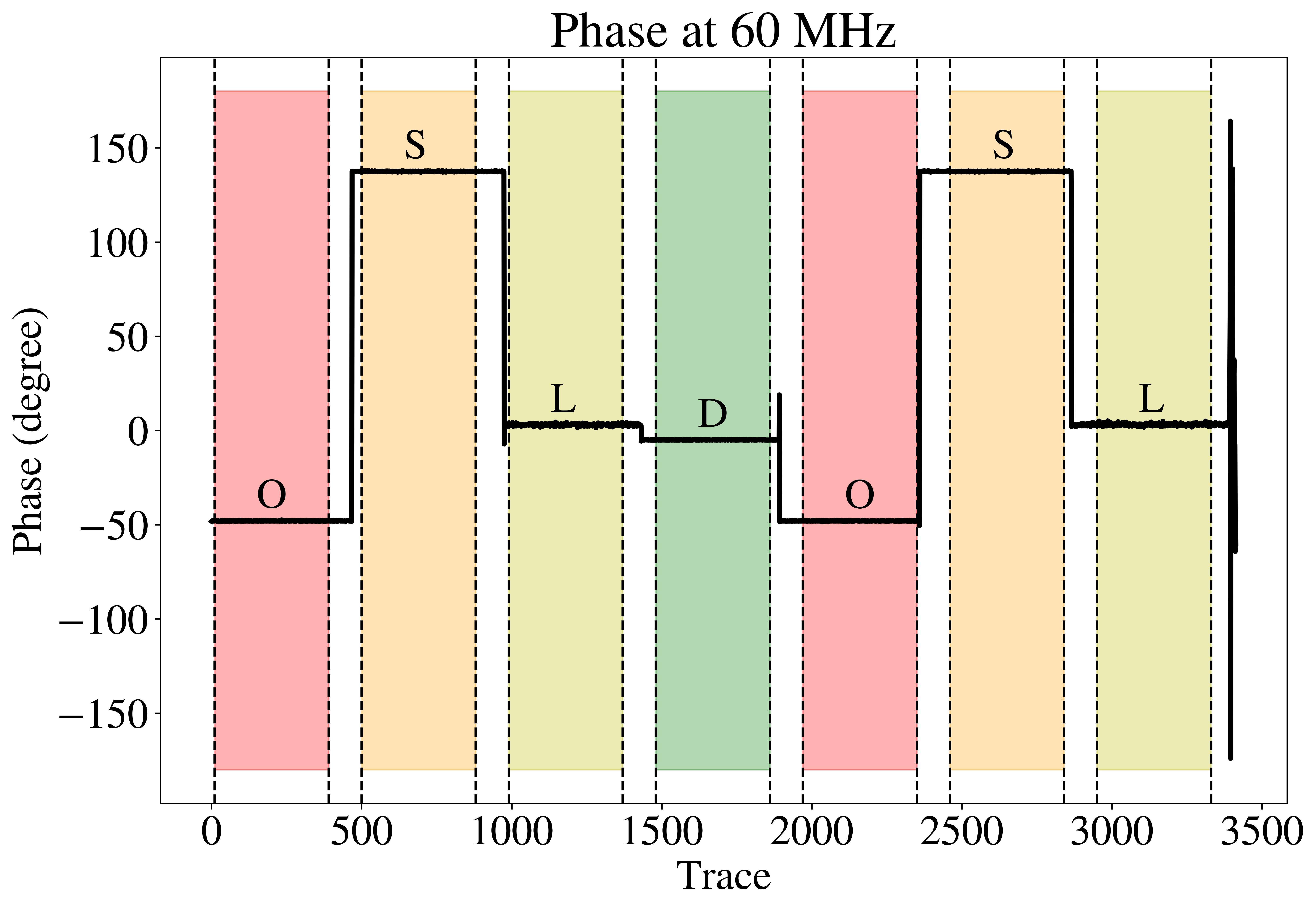}
    \caption{\textit{Top:} Waterfall images showing the reflected power and phase across different states. The x-axis shows the trace number, which represents time. The order of states is O-S-L-D-O-S-L, with 10 minutes for each state. \textit{Bottom:} The black line plot shows the power and phase at an arbitrary frequency of 60~MHz for the 7 states in the sequence (shaded distinctly for clarity).}
    \label{fig:states_visualisation}
\end{figure}

In Figure~\ref{fig:states_visualisation}, reflected power and the respective phase for a given calibration cycle are shown for the 7 different system states. 
The auto-correlation power measured for each state at ADC 1, corresponding to the RF path containing the terminations, is corrected for any variation in the noise source output using the respective simultaneous reference power measurements from ADC 2. 
The cross-correlation of the two ADC inputs (RF and reference paths) provides us with the phase measurement of each state with respect to the reference. We therefore acquire a complex referenced measurement for each termination. 
Finally, we average all traces for each termination to yield four referenced complex traces, corresponding to Open, Short, Load and DUT. The complex averaging of PSD measurements spread over time is expected, in some cases, to reduce gain drifts that may occur in the network downstream of the reference splitting at the coupler C1 in Figure~\ref{fig:s11_block}. Additionally, all measurements are performed only after a warm-up period of approximately 30~min, for the components to reach thermal equilibrium.

\begin{figure}[!htbp]
    \centering
    \includegraphics[width=0.8\linewidth]{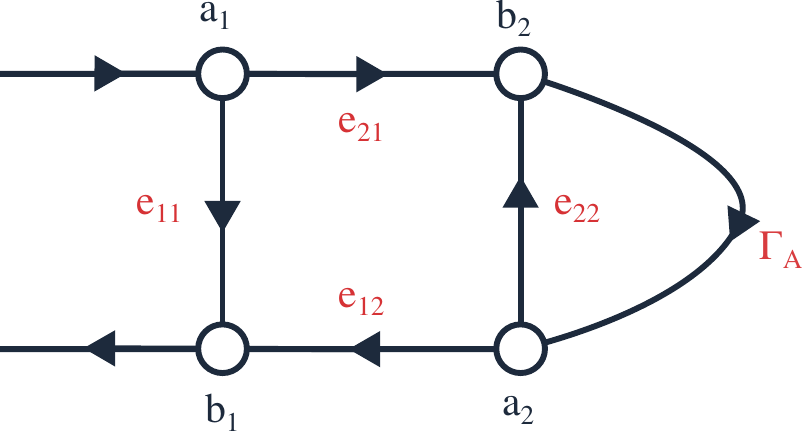}
    \caption{The signal flow diagram for our setup. Here the error terms are as follows: $e_{11}$ is directivity error, $e_{21}e_{12}$ is reflection tracking error and $e_{22}$ is source match error. $\Gamma_\mathrm{A}$ is the actual reflection coefficient of the device. Reflection coefficient $\Gamma_\mathrm{M}$ is measured at the left-hand side as $\frac{b_1}{a_1}$. The image is adapted from Rytting (1996) \cite{Rytting_error_model}.}
    \label{fig:agilent_1_port}
\end{figure}

The referenced complex measurements are used in the VNA one-port calibration, which is performed using the 3-term error-correction model \cite{Rehnmark_VNA_calibration_process, Rytting_error_model}.
This is explained using Figure~\ref{fig:agilent_1_port} with a signal flow diagram. For a given one-port setup, systematic errors between the components in Figure~\ref{fig:s11_block} are classified into three types:
(a) directivity error due to coupler leakage or directivity at C2 ($e_{11}$);
(b) reflection tracking error due to differences in response of ADC 1 and 2 ($e_{21}e_{12}$); and
(c) source match error due to impedance mismatch of the test port ($e_{22}$) \cite{VNA_error_calibration}.
The model here is simplified using network flow-graph rules to get a solution given by
\begin{equation}
    \Gamma_\mathrm{M} = e_{11} + \frac{(e_{21}e_{12})*\Gamma_\mathrm{A}}{1-e_{22}\Gamma_\mathrm{A}},
\label{one-port}
\end{equation}
where $\Gamma_\mathrm{M} (=\frac{b_1}{a_1})$ and $\Gamma_\mathrm{A}$ are the measured and actual reflection coefficients of the DUT.
Equation~\ref{one-port} can be rewritten as
\begin{equation}
    \Gamma_\mathrm{M} = e_{11} + (e_{21}e_{12} - e_{11}e_{22})\Gamma_\mathrm{A} + e_{22} \Gamma_\mathrm{M} \Gamma_\mathrm{A}, 
\end{equation}
and solved as a system of linear equations using three standard $\Gamma_\mathrm{M}$ measurements with known $\Gamma_\mathrm{A}$ values to get the error terms. In our exercise, the three known $\Gamma_\mathrm{A}$ values correspond to the ideal values of the three standard terminations.
Once derived, the error terms are used to correct the $\Gamma_\mathrm{M}$ of the DUT.

\subsection{Results and discussion on performance}

\begin{figure}[!htbp]
    \centering
    \includegraphics[width=1\linewidth]{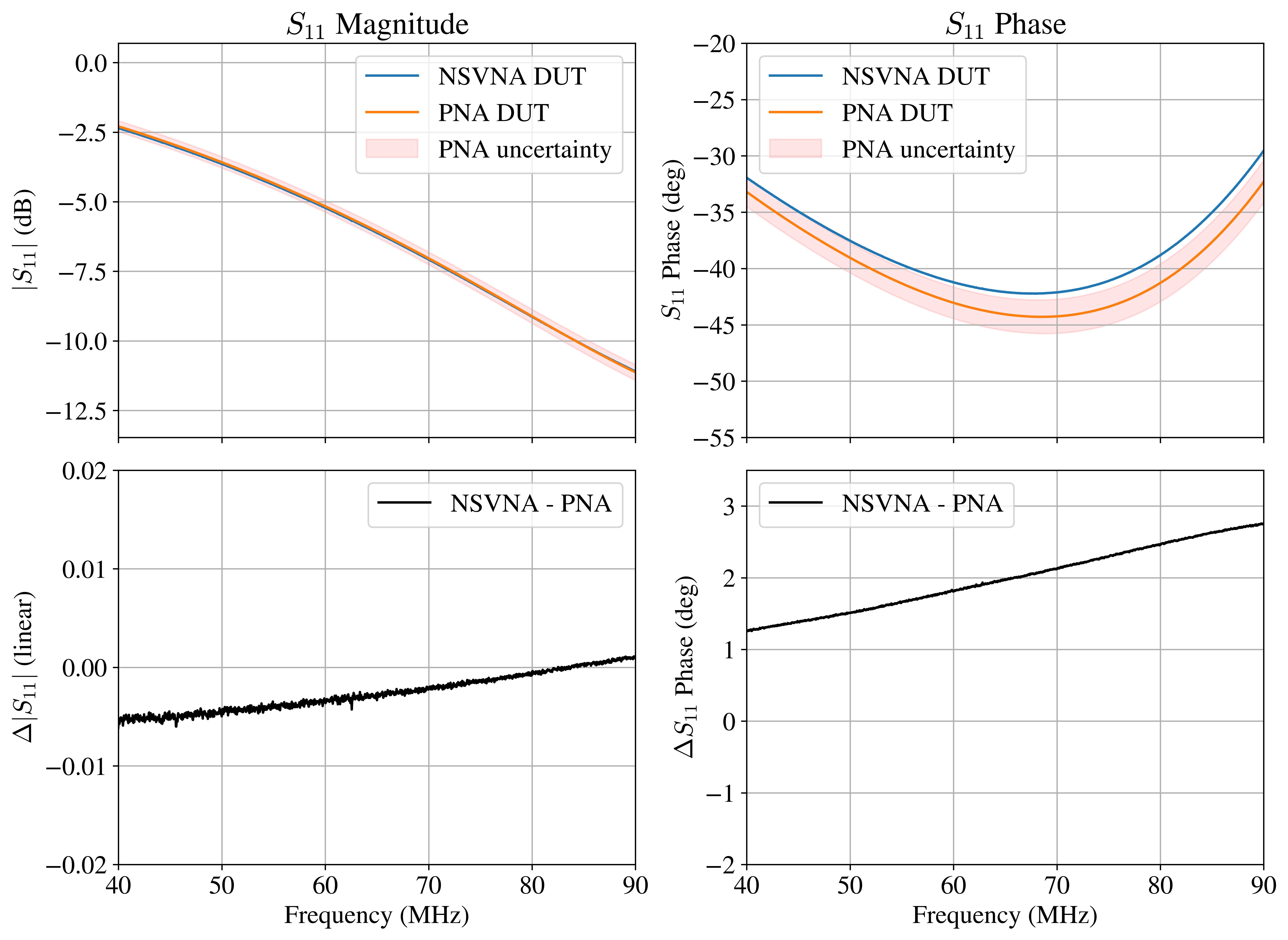}
    \caption{\textit{Top:} Plots showing the final calibrated $S_{11}$ (magnitude and phase) measured using the NSVNA and a comparison with the PNA measurement for the same DUT - the custom-made RLC circuit. The PNA calibration and measurement were performed with an averaging factor of 200. Shaded in red is the PNA uncertainty. \textit{Bottom:} Residual plots showing the difference between the NSVNA and PNA measurements in linear units and degrees for magnitude and phase measurements, respectively.}
    \label{fig:in_situ_calibrated}
\end{figure}

We test the setup by employing an RLC network as the DUT. The final calibrated $S_{11}$ of the RLC network is shown in the top panels of Figure~\ref{fig:in_situ_calibrated}. 
We note that the standard uncertainty\footnote{The standard uncertainty is calculated using the Type A analysis as $u_\mathrm{A} = \sigma/\sqrt{n}$, where $\sigma$ is the experimental standard deviation for $n$ measurements \cite{JCGMGUM}.} over a 5-cycle observation is close to $10^{-4}$ in magnitude and $0.01^\circ$ -- $0.04^\circ$ in phase, over the band.
Further, as a first-order test of the measurement accuracy, the $S_{11}$ of the same DUT was also measured using the Agilent Technologies PNA-X N5241A VNA and is overlaid with the NSVNA measurement in the same figure. Additionally, the PNA measurement error bar is also shown as a red shaded region calculated from the VNA uncertainty calculator (and as plotted in Figure~\ref{fig:PNA_calculated_uncert}).
The bottom panels of Figure~\ref{fig:in_situ_calibrated} show the difference between the two measurements. The differences range between 0.002 -- 0.005, which represents a good agreement between the two measurements. 
While the PNA measurement serves as a good accuracy test for the NSVNA unit, it is limited by the reflection uncertainties claimed for the device, which are of the order 0.01 and  $1^\circ$ in magnitude and phase, respectively \cite{PNA_datasheet}. For magnitude, the NSVNA and PNA measurements differ within these error bars.

We report a phase discrepancy of $1^\circ-3^\circ$ relative to the PNA measurements, which is significant in the context of the noise-wave calibration used in SARAS \cite{Meys_noise_wave, SARAS3paper}. The discrepancy was observed to be repeatable across multiple measurements; however, it proved difficult to remove in a subsequent correction procedure.
Several factors may contribute to this behaviour.
First, the calibration assumes ideal standard terminations. Although the calibration was repeated using polynomial models \cite{VNA_handbook} of the standards, no significant improvement in recovered phase was observed. 
Additionally, ageing of the standards could cause the open capacitance, short inductance, and load mismatch to deviate from the quoted values in their datasheet.
The second factor is the limited directivity of coupler C2 (ref. Table~\ref{tab:Additional_considerations}), which plays a critical role in separating the incident and reflected signals.
In combination with imperfect standards, this may lead to an inaccurate estimation of the VNA error terms, specifically the directivity ($e_{11}$).
Consequently, the model and choice of precision calibration standards, as well as the directivity of C2, represent important limitations of the current calibration approach \cite{Keysight_1port}.
Further, contributions may arise from impedance mismatches downstream of Amplifier~1, non-linearity of the amplifiers and receiver, reference plane mismatch between the four terminations, and differential phase and gain drifts and finite isolation between the RF and reference paths. These effects can be intensified by adapter imperfections and connector repeatability errors at the interface between the terminations and the switch, which serves as the VNA reference plane. 
A more detailed investigation of these effects will be undertaken in future iterations of the setup.

As part of this effort, we plan to investigate improved standards using data-based calibration kits, where the standard responses are provided as traceable data, which are determined through rigorous physical and electrical models, complemented by high-accuracy reference measurements \cite{Callegaro_2009}. Such standards are expected to reduce modelling errors associated with simplified third-order polynomial models used for the present work \cite{CalKit_intro}.
A similar method can be used for verification, where measurement of appropriately selected standards, chosen to cover a broad region of the Smith chart, can provide sufficient confidence that the NSVNA result is consistent with the traceability chain \cite{Traceability_Wong, NIST_VeridiCal}.
The calibration can be further improved by using multiple standards and weighted least squares methods based on the knowledge of the calibration standards \cite{WLS_calibration, Hoffman_2010_uncertS11}. 
Alongside, characterisation techniques for calibration standards can be explored, following approaches such as those in Monsalve et al. (2016) \cite{Monsalve_DR_50ohm_characterisation}, which introduces asymmetrical passive network measurements in direct/reverse modes for this purpose.

\section{Smoothness analysis} \label{Sec:result_validation}

In this section, we perform a spectral smoothness assessment of the NSVNA for 21-cm cosmology applications in the context of SARAS-like experiments. We begin by studying the spectral complexity of the measured DUT $S_{11}$ with theoretical expectations. Then we examine the impact of this on the spectral smoothness of sky measurements that could impact the 21-cm signal inference.

For the spectral smoothness analyses, an MS polynomial fit algorithm is used \cite{MaxSmooth, Harry_maxsmooth_EDGES}.
The model is parametrised as a polynomial in the logarithm of frequency. The fitted polynomial is then obtained by exponentiating the model on a linear scale. 
The fitting starts with a regular 2nd-order polynomial fit, whose coefficients serve as an initial guess for the MS fitting procedure. A loop then continues with specified inverse-variance weights and the allowed number of inflection points. We use uniform weights for all MS fits, except for sky measurements, where physically motivated thermal-noise weights are opted for. 
While the model is parametrised in logarithmic space, the optimisation minimises the chi-squared on a linear scale. 
The loop progresses with increasing polynomial degrees, using the solutions from each iteration as guess values in the next one. The process terminates when the root mean square saturates. 
The arbitrarily high-order polynomial fits in these cases are allowed, considering that the constrained nature of the polynomial would not fit out the signal of interest, i.e. the global 21-cm signal, in the later analyses \cite{Rao_2017}. However, if the number of inflections allowed is increased, the MS polynomials would fit a greater fraction of the signal, thus reducing the sensitivity to the signal.

\subsection{Spectral analysis and analytical model for the RLC} \label{subsec:result_validation_analytical} 

The first-order RLC network was designed to be maximally smooth with at most one inflection point, avoiding multiple zero crossings in its higher-order derivatives \cite{MaxSmooth}. This was done to replicate the behaviour of the SARAS antenna over a frequency range of 40-90~MHz, where such a spectral nature of $S_{11}$ helps in separating foregrounds from the 21-cm signal. For all analyses, we restrict the bandwidth to this range, which is set by the SARAS receiver smoothness specification \cite{SARAS3paper}. 

\begin{figure}[!htbp]
    \centering
    \includegraphics[width=0.95\linewidth]{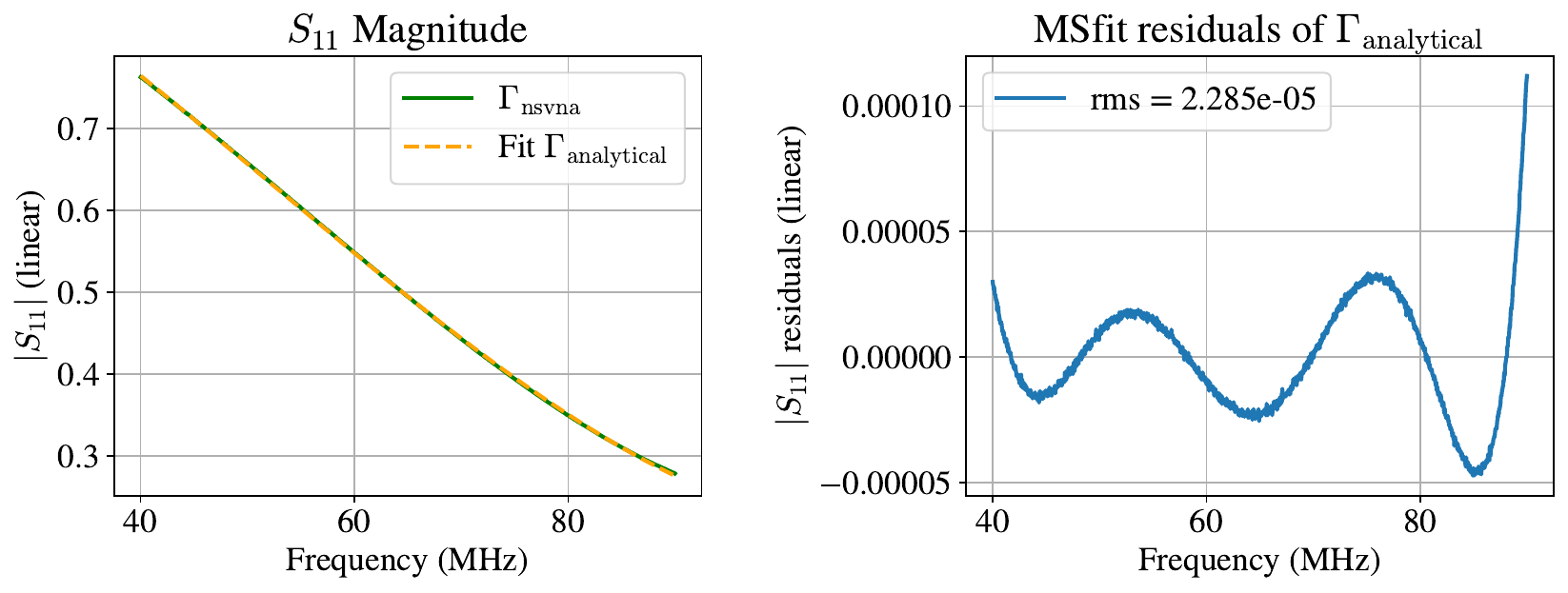}
    \caption{\textit{Left:} Analytical fit to the RLC $\absval{S_{11}}$ measured using the NSVNA. \textit{Right:} Residuals after fitting a $9^\mathrm{th}$ order MS function allowing one inflection point to $\absval{\Gamma_\mathrm{analytical}}$ for the RLC. It yields residuals of rms $\sim2.285 \times 10^{-5}$.}
    \label{fig:rlc_process}
\end{figure}

First, we estimate the intrinsic smoothness of the first-order RLC network. For this purpose, we modelled the reflection coefficient of the RLC circuit along with the parasitic impedance in the circuit using a lumped-element model. The model takes in the actual component values as the initial guess. It then determines the R, L, C and l (length) of the lumped circuit that produce the best fit to the $S_{11}$ measured with the NSVNA. The fit metric was the rms of the fit residuals.
The optimised R, L, C and l values were used to construct an analytical form for $S_{11}$ of the RLC device. The smoothness of this analytical $S_{11}$ was evaluated by fitting an MS function to its magnitude after adding to it a subdominant Gaussian noise with a standard deviation of $10^{-6}$.
The rms of the residuals for this fit was approximately $2.285 \times 10^{-5}$. The analytical form of $\absval{S_{11}}$, and its MS fit residuals, are shown in Figure~\ref{fig:rlc_process}. This result sets a best-case metric for the smoothness of the RLC circuit. To avoid possible circularity in our assessment, a similar exercise was repeated with PNA measurements as template, yielding MS fit residuals of rms $\sim 5.819 \times 10^{-5}$. Although the NSVNA-derived analytical $S_{11}$ shall be considered primarily for all tests, we shall quote corresponding results for PNA-derived analytical $S_{11}$ as well.


\begin{figure}[!htbp]
    \centering
    \includegraphics[width=0.95\linewidth]{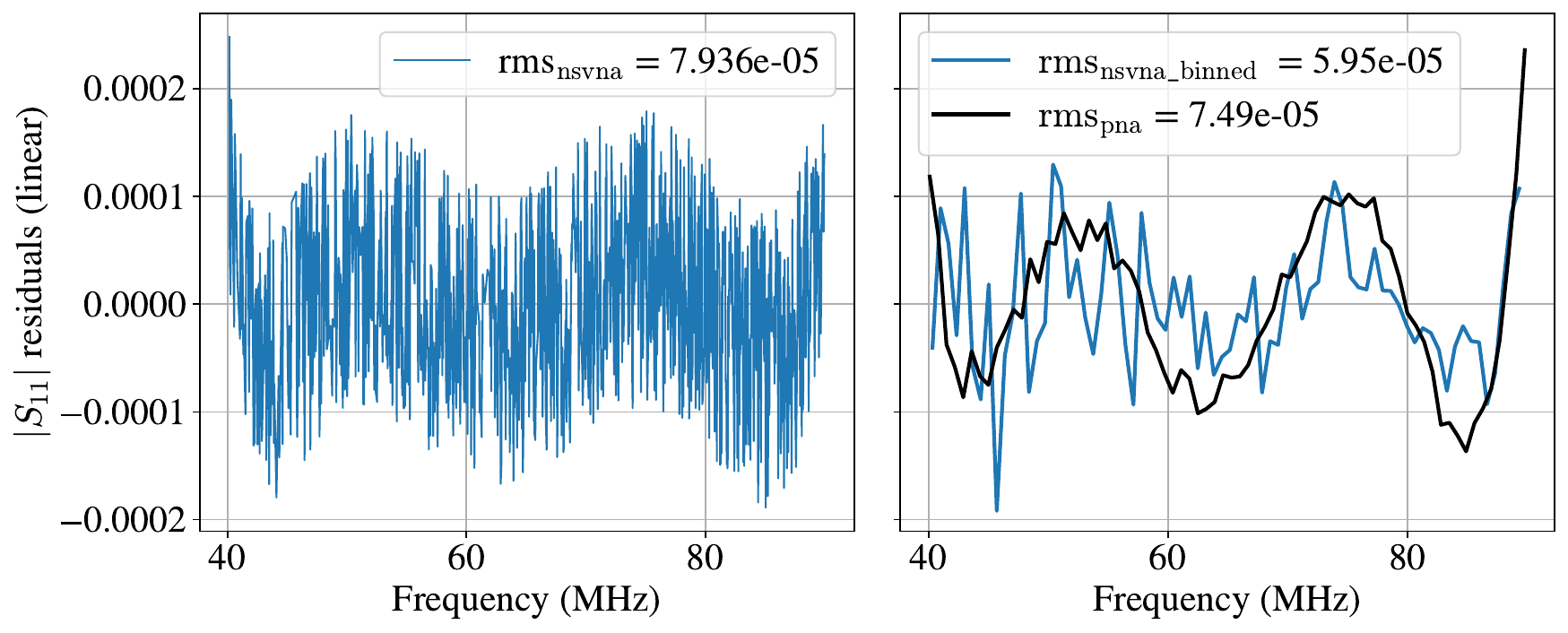}
    \caption{\textit{Left:} The residuals after fitting a $9^\mathrm{th}$ order MS Function allowing one inflection point to magnitude of $S_{11}$ measured using the NSVNA. The rms of these residuals is about $7.936 \times 10^{-5}$. \textit{Right:} A comparison plot between PNA and NSVNA residuals, where the NSVNA residuals (of the left plot) are binned to match the frequency resolution of the PNA measurement closely.}
    \label{fig:insitu_s11residue}
\end{figure}

Following this estimate, fitting an MS function with one allowed inflection to the NSVNA-measured $\absval{S_{11}}$ would be a reasonable metric to ascertain the level of non-smooth systematics in the measurement. Due to the unshielded condition of the measurement, an RFI flagging preceded this fitting, flagging 591 out of 1639 channels. Fitting a $9^\mathrm{th}$ order MS function to the RFI-flagged $\absval{S_{11}}$ yielded a residual of rms of $7.936~\times~10^{-5}$, which is presented in the left panel of Figure~\ref{fig:insitu_s11residue}. The order of these residuals is usually consistent after the warm-up period of the NSVNA. 
We observed that the residual rms values for both the NSVNA measurement and the analytical model are of the same order. Additionally, the noise-like spectrum of the residuals suggests that the measurement is noise-limited. The right panel of the figure shows a comparison with MS fit residuals of the PNA measurement, where the NSVNA residuals are binned to match the PNA frequency resolution. The binned NSVNA data were observed to be structurally smoother (rms~$\sim6\times10^{-5}$) than the other.

\subsection{Impact on sky measurements} \label{Sec:sky_impact}

The objective of developing the \textit{in situ} device is to enhance the calibration of the radiometer and, consequently, improve the detectability of the global 21-cm signal. For this purpose, we present a simplified assessment of the effects of the spectral nature of the reflection coefficients on sky measurement data in this section.

The sky model is chosen from Harker (2015) \cite{Harker_foreground}, where the author has assumed an antenna with a flat response, observing a quiet region of the global sky model \cite{GSM_de_Oliveira}. The sky temperature convolved with the antenna beam is given by
\begin{equation}
    \ln{T_\mathrm{ant}} = \ln{T_0} + \sum_{i=1}^{3}a_i [\ln{(\nu/\nu_0)}]^i ,
    \label{eq:fg}
\end{equation}
where $\nu_0$ = 80~MHz and $\{T_0, a_1, a_2, a_3\}$ = $\{2039.611, -2.42096, -0.08062, 0.02898\}$~K. Additionally, we add a subdominant mK-level Gaussian random radiometric noise and assume a lossless antenna in our analysis. 
The component of antenna temperature measured after including the effect of antenna mismatch is calculated as
\begin{equation} \label{eq:T_obs}
    T_\mathrm{obs} = T_\mathrm{ant} (1-\absval{\Gamma_\mathrm{true}}^2) \equiv T_\mathrm{ant} (1-\absval{\Gamma_\mathrm{analytical}}^2),
\end{equation}
where the final substitution comes from the discussion in Section~\ref{subsec:result_validation_analytical}, which treats the analytical model of the RLC circuit as a limit on the smoothness that can be obtained from its $S_{11}$ measurement. 
This equation further assumes a perfectly matched LNA at the next stage.

\begin{figure}[!htbp] 
    \centering
    \includegraphics[width=0.95\linewidth]{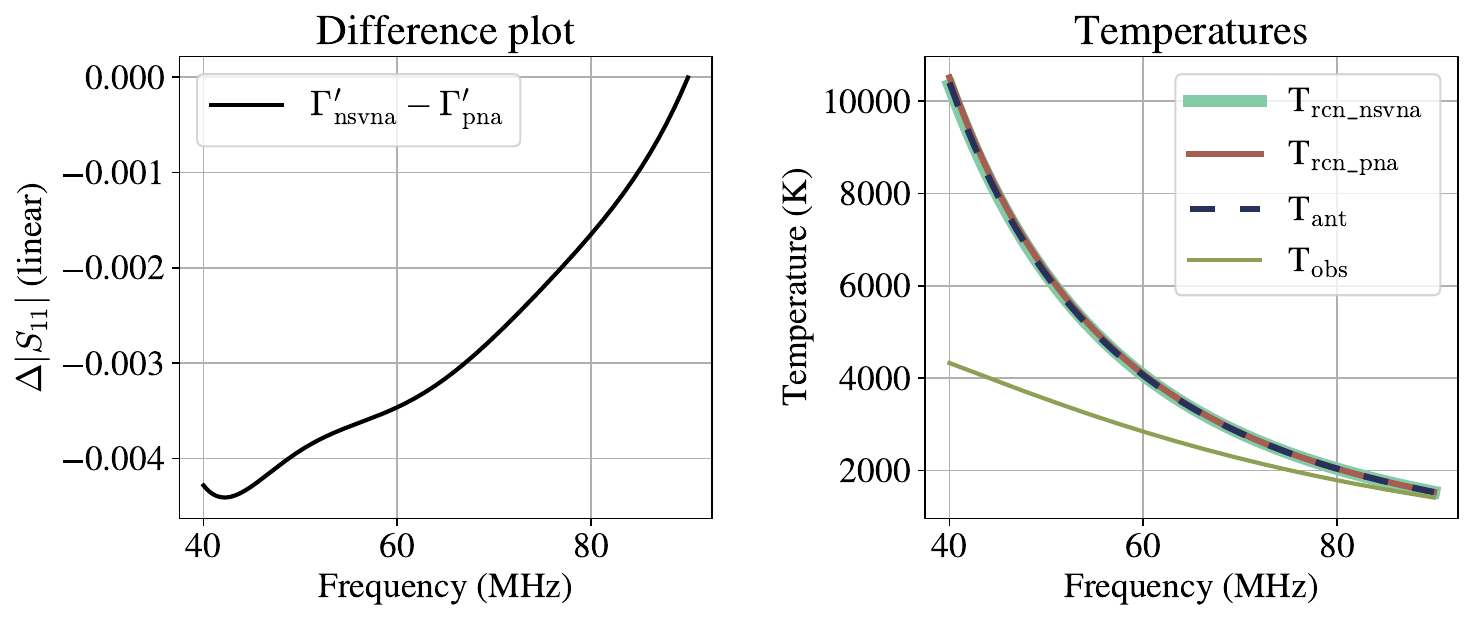}
    \caption{\textit{Left:} A difference plot between the modelled reflection coefficients.
    \textit{Right:} The true antenna temperature (Equation~\ref{eq:fg}) is shown as a dashed curve. This is plotted over the reconstructed antenna temperatures (Equation~\ref{eq:T_reconstructed}) derived using MS modelled $\Gamma'_\mathrm{nsvna}$ and $\Gamma'_\mathrm{pna}$. 
    The observed antenna temperature (Equation~\ref{eq:T_obs}) lies at a lower level than $T_\mathrm{ant}$, illustrating the power loss due to antenna mismatch.}
    \label{fig:Gamma_Tsky}
\end{figure}

A direct inversion of Equation~\ref{eq:T_obs} would give us the true antenna temperature after the beam-convolution in Equation~\ref{eq:fg}. To understand the effects of the measured reflection coefficients on the spectral smoothness of the data, we define reconstructed antenna temperatures, where we `correct' the measured sky spectrum, $T_\mathrm{obs}$. 
These are constructed using the reflection coefficients, $\Gamma'_\mathrm{nsvna}$ and $\Gamma'_\mathrm{pna}$, which are the corresponding modelled MS functions with one inflection, as expected for a first-order RLC network and argued in Section~\ref{subsec:result_validation_analytical}. It must be noted that, as this analysis makes use of the functional forms of the reflection coefficient, no RFI flagging was performed in this section. With these MS-modelled reflection coefficients, the reconstructed antenna temperatures are defined as  
\begin{equation} \label{eq:T_reconstructed}
\begin{split}
    T_\mathrm{rcn\_nsvna} &= T_\mathrm{obs}/(1-\absval{\Gamma'_\mathrm{nsvna}}^2) = T_\mathrm{ant}\frac{(1-\absval{\Gamma_\mathrm{true}}^2)}{(1-\absval{\Gamma'_\mathrm{nsvna}}^2)} , \\
    T_\mathrm{rcn\_pna} &= T_\mathrm{obs}/(1-\absval{\Gamma'_\mathrm{pna}}^2) = T_\mathrm{ant}\frac{(1-\absval{\Gamma_\mathrm{true}}^2)}{(1-\absval{\Gamma'_\mathrm{pna}}^2)} .
\end{split}
\end{equation}
A comparison of the two modelled reflection coefficients and the respective reconstructed antenna temperatures is shown in Figure~\ref{fig:Gamma_Tsky}. For reference, $T_\mathrm{obs}$ is also shown.  

\begin{figure}[!htbp]
    \centering
    \includegraphics[width=0.95\linewidth]{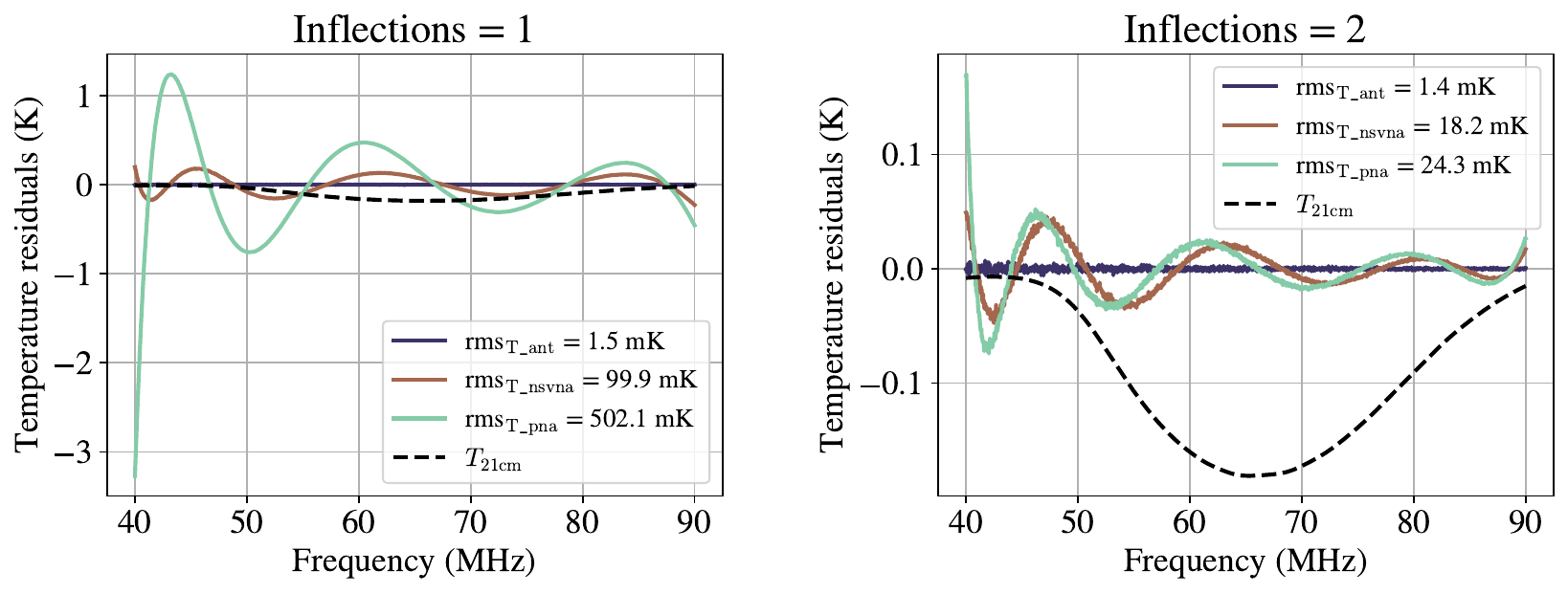}
    \caption{MS fit residuals of the reconstructed antenna temperatures, allowing one (left) and two (right) inflection points in the MS fit. Additionally, residuals for an MS fit to true antenna temperature $T_\mathrm{ant}$ are also shown. For reference, we also plot the standard $T_\mathrm{21cm}$ signal, used in the signal recovery test.}
    \label{fig:Tsky_msfits}
\end{figure}

As the final step, the smoothness of all three sky temperatures (the true $T_\mathrm{ant}$ and reconstructed $T_\mathrm{rcn\_nsvna}$ and $T_\mathrm{rcn\_pna}$) is determined using an MS fit. As the reflection coefficients are fitted using an MS function with one inflection point in the band, the reconstructed temperature MS fits are allowed to have inflections greater than or equal to one. It was observed that the rms of the fit residuals did not change significantly when increasing the number of inflections beyond two. The MS fit analysis was thus restricted to one and two inflections, the residuals of which are shown in Figure~\ref{fig:Tsky_msfits}.

For each allowed inflection, we find that the reconstructed antenna temperature using the NSVNA measurement ($T_\mathrm{rcn\_nsvna}$) yields lower residuals after subtracting the MS fit compared to that for the one using the PNA measurement ($T_\mathrm{rcn\_pna}$). This difference is more prominent in the one-inflection case. Expectedly, both these cases result in higher rms compared to the true antenna case, where the reflection coefficient is assumed to be perfectly corrected for. This clearly indicates that the nature of non-smooth systematics in PNA is higher compared to the NSVNA method, albeit consistent with its measurement accuracy. 
To ensure that the results were not influenced by the choice of the true reflection-coefficient template, the analysis was repeated using the analytical reflection coefficients derived from the PNA measurements in Section~\ref{subsec:result_validation_analytical}.
For the one-inflection case, the MS fit yielded residuals of 100~mK and 572~mK for $T_\mathrm{rcn\_nsvna}$ and $T_\mathrm{rcn\_pna}$, respectively. For two inflections, the corresponding residuals were 19~mK and 21~mK.

We now aim to translate these calibration residuals to their impact on global 21-cm signal detection. Any assessment of this kind needs an assumption on the spectral profile of the signal \cite{Agrawal_Direction_dependent_effects_2024}. Therefore, we assume one of the standard signal profiles, shown in Figure~\ref{fig:Tsky_msfits}, as an illustrative example, which is injected into the reconstructed antenna temperature spectrum. We adopt a maximally smooth function with 2 inflections, referred to as the foreground model, representing intrinsic foregrounds along with calibration systematics associated with reflection coefficient measurements, as described above.

In the first case, the spectrum is fit to just the foreground model ($M_1$), while in the second case, the fit is that of the foreground model plus 21-cm signal ($M_2$). In both cases, we compute the $\chi^2$, defined by: 
\begin{equation}
    \chi_j^2 = \sum_{i=1}^N\frac{(d_i - M_{ij})}{\sigma_i^2},
\end{equation}
where $d_i$ is the data at the $i^{th}$ frequency, $M_{ij}$ is the $j^{th}$ model ($M_1$ or $M_2$) at the $i^{th}$ frequency, $\sigma_i$ represents the statistical uncertainty at the corresponding frequency, with measurements spanning over $N$ frequencies.

For signal detection, $\chi^2$ should yield a lower value for $M_2$ than $M_1$, as also noted in \cite{0004-637X-815-2-88}. The significance in reduction of $\chi^2$ is assessed by using Bayesian Information Criterion (BIC) \cite{c4048c8f-6ca9-3965-96a3-653ab8996955}, given by:
\begin{equation}
    \mathrm{BIC} = k \ln (n) - 2 \ln (\hat{L}),
\end{equation}
where $\hat{L}$ is maximum likelihood value, assuming Gaussian distribution, $n$ is the number of parameters, and $k$ represents the total number of model parameters. The difference in BIC, $\Delta \mathrm{BIC}$, between models $M_2$ and $M_1$, therefore, indicates the significance of detection. We note strong detections in all cases: $T_\mathrm{ant}$, $T_\mathrm{rcn\_nsvna}$ and $T_\mathrm{rcn\_pna}$, with $\Delta \mathrm{BIC} \sim 10^4$. 

However, we note the different systematic floors for NSVNA and PNA cases, manifested as different rms residuals, as shown in Figure~\ref{fig:Tsky_msfits}. This may limit a robust detection of global 21-cm signals that may be of lower amplitudes or partially located in the band of interest \cite{Cohen_Fialkov_21cm_parameter_space}; the limit is primarily decided by the respective systematic floors. A wide-scale parametric exploration of global 21-cm astrophysical space, and the associated constraining power with varying levels of residual systematics, will be covered in future work. Nevertheless, given the miniaturised, field-rugged development of NSVNA, with focus on its ease of integration with radiometers, the lower systematic floor and the signal injection test show its suitability for global 21-cm experiments, such as SARAS and PRATUSH. 
These results, however, require assessing the impact of additional instrumental effects relevant to a complete radiometer measurement, such as beam chromaticity, antenna radiation efficiency, impedance mismatch effects, $S_{11}$ phase measurements in noise-wave calibration, and so on, which clearly require stand-alone focused investigation.

\section{Summary} \label{Sec:Summary}
The detection of the global 21-cm signal relies strongly on the calibration of the radiometer. In this paper, we focused on one aspect of this calibration, that is, the non-smooth spectral structures in the measurement of the antenna return loss. This is essential as the return loss directly imprints features in the measured sky spectrum as well as the receiver noise waves. We discussed the novel design and development of an \textit{in situ} noise source-based VNA (NSVNA) for the SARAS radiometer, contrasting its level of non-smooth calibration systematics with a high-accuracy off-the-shelf VNA, the PNA-X N5241A 423/029.

The developed NSVNA utilises a noise source and employs a cross-correlation spectrometer to measure the magnitude and phase of the reflection coefficient. Notably, the noise source-based architecture enabled simultaneous frequency measurements across the band, and the dual auto-correlation ensured simultaneous measurement of reflected and reference signals, making the process immune to gain variations. Moreover, this method utilises minimal components specifically designed for measuring the antenna's reflection coefficient and can be considered an advancement over previous methods. With this design and calibration methodology, we aimed to reduce all three kinds of measurement errors in the VNA setup, namely systematic, random and drift errors. 

We presented the $S_{11}$ measurement results for a test RLC circuit and reported a standard uncertainty of the order of $10^{-4}$ and $0.01^\circ$ -- $0.04^\circ$ in linear magnitude and phase, respectively, for the NSVNA. Later, we discussed two tests to qualify the designed VNA, specifically assessing the spectral characteristics of the measurement. As a first level of validation, we compared the NSVNA measurements for the RLC circuit with those obtained using the PNA. While the difference in phase measurement is in the order of $1^\circ - 3^\circ$, close to that of the reported PNA uncertainty ($\sim1^\circ$), there is an excellent match in magnitude with residuals close to $10^{-3}$, well within the typical PNA uncertainty ($\sim0.01$).
For further qualification tests, we implemented the condition of maximal smoothness for the test RLC circuit. We begin by testing the theoretical smoothness of the circuit and find it to be with rms~$\sim~2.285~\times~10^{-5}$. Fitting an MS function to the $S_{11}$ magnitude measurement, we get residuals of rms close to $7.936 \times 10^{-5}$, suggesting the noise-limited nature of the NSVNA measurement. 
For the given test configuration and the RLC, the NSVNA measurements yielded marginally lower non-smooth calibration residuals than the PNA.

As an illustrative test of the work, the effects of the NSVNA and PNA measurements were evaluated on mock sky measurements. 
These tests were performed treating an analytical model as the template, which was derived using the NSVNA measurements. The NSVNA-reconstructed antenna temperatures yielded five times lower levels of non-smooth systematics than those from PNA, with an rms of approximately 100~mK when allowed one inflection across the band. The NSVNA performed marginally better for the two-inflection case as well. When the PNA data were used to optimise the analytical model, similar results were observed.
This hinted at the presence of non-smooth systematics in PNA calibration, given the calibration configuration and advocated the importance of specialised reflection coefficient measurement systems for precision cosmology experiments.
Finally, we performed the signal recovery test using a standard 21-cm template, where the significance of recovery was evaluated using the BIC.
We observed that all cases ($T_\mathrm{ant}$, $T_\mathrm{rcn\_nsvna}$ and $T_\mathrm{rcn\_pna}$) favoured a signal detection with similar significance. However, we noted that a higher systematic floor for the PNA case may limit the sensitivity to signals with lower amplitudes or those that lie only partially within the observed frequency band.

We emphasise that measurements made in the lab using a PNA cannot be used for corrections in the field because of drifts and electromagnetic effects that can occur due to a change in the environment of the radiometer. Moreover, PNAs are bulky and not typically field-rugged, which limits their field applicability for global 21-cm experiments due to difficulties in integrating with the system for \textit{in situ} measurements.
The noise-source-based method provides an alternative for such measurements. Its ease of integration with SARAS and PRATUSH systems, together with the validation tests, demonstrates its potential for application in global 21-cm experiments.
We plan to work further on miniaturising and integrating this device with the SARAS receiver on an integrated PCB circuit. This is expected to improve the spectral quality by providing better shielding against RFI, as well as lower loss and higher directivity where required in the signal chain. The choice of surface-mount devices may require additional changes to the circuit to preserve (or improve) the accuracy and spectral smoothness of the measurements. This has to be done based on the available surface-mount devices with better (or similar) characteristics like low loss, good directivity and isolation in the band of our interest.
Simultaneously, future work involves a thorough characterisation of the NSVNA accuracy and a comprehensive assessment of the measurement effects, which involves other elements in the receiver model.
Following this design philosophy, an \textit{in situ} noise source-based VNA can eventually benefit space-based experiments like PRATUSH, where the need for an \textit{in situ} $S_{11}$ measurement becomes more crucial.

\newpage

\backmatter





\bmhead*{Acknowledgements}

The authors thank the members of the CMB Distortion lab at Raman Research Institute for providing the resources and support necessary for this project. 
We also thank Kasturi S. for the design and development of the filters and RLC network, and the Electronics Engineering Group (RRI) for the test and design of several components of the system. We extend our thanks to the Mechanical Engineering Group (RRI) for the construction of the chassis and support structures for the system.
The authors express gratitude to Tejas Oak, Yogen Pranesh and Kinjal Roy for the constant support and countless `problem-solving' sessions during the project.
Finally, AKD thanks Rohan Patel at Cavendish Laboratory, University of Cambridge, for the valuable discussions on VNA calibration methods and error models.

\section*{Declarations}

\textbf{Funding}

Not applicable.

\noindent \textbf{Conflict of Interest}

The authors have no relevant financial or non-financial interests to disclose.

\noindent \textbf{Ethics approval and consent to participate}

Not applicable.

\noindent \textbf{Consent for publication}

Not applicable.

\noindent \textbf{Data Availability}

Data sets generated during the current study are available from the corresponding author on reasonable request.

\noindent \textbf{Materials availability}

Not applicable.

\noindent \textbf{Code availability}

Not applicable.

\noindent \textbf{Author contribution}

AKD led the writing of the manuscript. SR and AKD conceived the novel noise source-based design of the RF and Optical Fibre system as well as its optimisation and system implementation. AKD led the system-level testing and verification of the VNA. 
YA worked on the investigation of the impact of NSVNA on sky measurements and 21-cm inference. JAT led the project for miniaturisation and integration of the instrument with SARAS and PRATUSH receivers. 
VG, KS and SR worked on earlier iterations of the design, leading to the final output. AKD and SS developed techniques for validation of the system. 
SS and MSR provided supervisory support throughout the work and editorial advice on the manuscript. SKS and BSG led the digital spectrometer design, essential for the complex reflection coefficient measurement.

\bibliography{sn-bibliography}

\end{document}